# Sleep and redistribution preferences: Considering allowable tax rates

Eiji YAMAMURA

Department of Economics, Seinan Gakuin University/ 6-2-92 Nishijin Sawaraku Fukuoka, 814-8511.

Email: yamaei@seinan-gu.ac.jp

Fumio OHTAKE

Center for Infectious Disease Education and Research, Osaka University, Japan,

Email: ohtake@cider.osaka-u.ac.jp

# Abstract

This study explored the association between sleep duration and redistribution preferences. Using an online survey, we propose a hypothetical situation in which the tax paid directly by respondents is redistributed to those earning less than one-fifth of the respondents' income. Next, we asked about the allowable tax rates. We found the following through Tobit and ordered logit regression estimations. (1) The relationship between sleep hours and the allowable tax rate showed an inverted U-shape, where the optimal amount of sleep led to the highest allowable tax rate. (2) High-quality sleep was more positively correlated with the allowable tax rate than was low-quality sleep when the sleep quantity was the same. (3) Sleep hours were more significantly and positively correlated with the allowable tax rate in the high-income group than in the low-income group. (4) Assuming that twice the amount of tax paid goes to those with lower income, individuals who previously preferred a higher tax rate were more likely to increase the allowable tax rate.

JEL Classification: H23, D63, D64

Key words: Sleep hours, Allowable tax rate, Preferences for redistribution, Altruism

# 1 Introduction

Sleep duration is widely acknowledged to have a detrimental effect on individuals.[1] As sleep duration shortens, people may be more self-interested, which means they may not consider others or society. Sleep is possibly related to tax issues, such as tax evasion or the perceived tax burden. However, no study has connected sleep to public financing. Primarily, we consider the correlation between sleep and subjective views on taxes. Furthermore, we explore how altruism is associated with the correlation between sleep and the allowable tax rate. While this method is very simple, it is not sufficiently rigorous to show causality. At the same time, the study also provides a new perspective that expands the scope of public finance and economics.

Sleep deprivation leads individuals to be aggressive, increasing the probability of violent incidents (Kamphuis et al., 2012; Wang et al., 2016). A lack of sleep makes it difficult to control one's emotions, resulting in anger, hostility, and aggression (Demichelis et al., 2023). Sleep deprivation was pervasive during the COVID-19 pandemic (Cellini et al., 2020, 2021; Di Giorgio et al., 2021; Jahrami et al., 2021). This may have increased tension between people, triggering conflicts in some cases. Lack of sleep increases aggression in different social groups (Saghir et al 2018) and poor sleep quality is strongly associated with poverty (Patel et al. 2010). Hence, low-income earners foster resentment toward high-income earners (Buttrick and Oishi, 2017, Gordon, and Chen, 2013). Redistribution policies were critical in mitigating the widening income inequality during the pandemic (Almeida et al., 2021).

---

[1] Sufficient sleep improves workers' productivity (Costa-Font et al., 2024; Gibson and Shrader, 2018; Kajitani, 2021; Monaco et al., 2005). As for academic performance, an increase in sleep duration is positively associated with educational attainment (Sabia et al., 2017). Short afternoon naps in the workplace significantly increase productivity and result in better subjective well-being (Bessone et al., 2021).

Researchers have found evidence that people become more prosocial with an increase in sleep duration and depth (Studler et al., 2024). Lack of sleep is negatively associated with trust, altruism, and helping behaviors (Anderson and Dickinson, 2010; Ben Simon et al., 2020; Dickinson and McElroy, 2017).

There are two possible reasons why insufficient sleep is negatively correlated with prosocial behaviors. First, sleep deprivation has a detrimental effect on social cognition abilities (Ben Simon and Walker, 2018; Guadagni et al., 2014, 2017), so the desire to help others decreases under conditions of sleep loss (Simon et al., 2022). Second, a lack of sleep hampers self-control and deliberative thinking (Anderson and Dickinson, 2010; Holbein et al., 2019). These functions are critical to enhancing prosocial behavior (Wyss and Knoch, 2022).

In this study, we hypothesize that sleep is positively associated with a preference for income redistribution to reduce income inequality.[2] Individuals were assumed to pay taxes for two reasons. First, people are motivated to contribute to reducing income inequality. This motivation is known as "altruism." Second, an "impure motivation" such as a "warm glow" means that people get some private benefit from their contribution

---

[2] In the field of economics, previous studies examining the preference for redistribution used discrete ordered variables (Likert scale) with redistribution policy as a proxy for redistribution preference(Alesina and Giuliano, 2009; Alesina and La Ferrara, 2005; Corneo and Grüner, 2002; Ohtake and Tomioka, 2004; Sabatini et al., 2020; Yamamura, 2012, 2014, 2015).

(Andreoni, 1989, 1990) [3].

We examined to what extent the sleeping hours of individuals was associated with the allowable taxes for income distribution to the poor. Using independently collected data, we found that (1) Sleep duration was positively associated with the allowable tax rate. (2) The association was strengthened if sleep quality improved. (3) In another hypothetical setting, in which we doubled the amount of tax distributed to people with lower incomes, the respondents were more likely to raise the allowable tax rate. (4) An estimation using a sub-sample showed that the positive effect of sleep duration on allowable tax rate was observed for the high-income group.

## 2  Data

### 2.1 Data Collection

Beginning in 2016, we constructed panel data using a comprehensive long-term survey of life and worldviews. The survey was independently designed.[4] Items related to the COVID-19 pandemic were inserted from the 2020 survey on. The title of the overall project is "Living Environment during Childhood and Current Life Attitudes," and we added the sub-title "A Study on the Influence of the New Type of Coronavirus Infection on Lifestyle Consciousness." The data used in this study were collected in February 2024, when the COVID-19 pandemic was almost over. Considering the situation in society, new questions on sleep were inserted into the 2024 survey.

---

[3] In the questionnaire used in this paper, individuals were asked about the allowable tax rate, i.e., the highest tax rate they were willing to accept.

[4] The primary objective of the survey was to explore broad socio-economic conditions rather than a single specific analysis. Although the survey is comprehensive in nature, it purposefully includes specific items designed to facilitate the analysis conducted in this study.

A research firm, the Nikkei Research Company (NRC), was commissioned to conduct the survey due to its significant experience with academic surveys. The NRC recruited panelists to register as reserve participants in the online survey. Panelists were asked to provide basic demographic information for a profiling database upon enrolment in the panel. A questionnaire was sent to panelists at the beginning of the survey. Once the target sample size was met, or slightly exceeded, the survey was closed to new panelists. To ensure representativeness, the sample was approximated to the population composition of Japan as a whole. For this purpose, respondents were collected from all over Japan covering ages 20–73. As it was an online survey, the sample was limited to internet users. Also, it should be noted that certain age groups were not included in the sample. To avoid ethical problems, those who were not adults were excluded. Individuals over 74 were not included because of possible cognitive limitations.

In total, 5,133 observations were gathered using the survey. However, some participants did not answer all questions related to the variables used in this study. Accordingly, the number of observations dropped to 3,731. Further, outliers or wrong answers were excluded. Finally, 3,476 observations were included in the sample for estimation. Further, observations were 1,625 and 1,851 for the high- and low-income groups, respectively.[5] Individuals with high-quality sleep were defined as those who went

---

[5] In the questionnaire, respondents were asked:

"Approximately how much was your total household income? Please choose one from the following choices. "

to bed between 21:00 and 24:00. The others were defined as having low-quality sleep. For robustness check, we also use other criteria when conducting the estimations.

## 2.2 Empirical Model

To test the impact of redistribution efficiency ($k$) on the optimal tax rate ($t$), we simply perform comparative statics. [6] In this study, we set the "Utility function" as below.

$$U = w \ln(Yp + ktY) + z(tY) + \ln(Y - tY) \qquad (1)$$

$U$ is utility. $Y$ is individual's own income level and $Yp$ is the aggregate income of others in society.

The first term, $w \ln(Yp + ktY)$, represents the individual's altruistic utility. By including $Yp$, the model assumes that the individual considers the overall social income level as a baseline for their altruistic concern. The term $ktY$ represents the actual resources reaching the poor as a result of the respondent's tax payment, amplified by the efficiency

---

1. Less than 1 million yen, 2. 1 to less than 2 million yen, 3. 2 to less than 4 million yen, 4. 4 to less than 6 million yen, 5. 6 to less than 8 million yen, 6. 8 to less than 10 million yen, 7. 10 to 12 million yen, 8. 12 million yen to less than 14 million yen, 9. 14 million yen to less than 16 million yen, 10. 16 to 18 million yen, 11. 18 to less than 20 million yen, 12. More than 20 million yen, 13. Do not want to answer

Midpoint of each category was used as the income level. Upper and lower bounds were converted to 50 and 2300 million yen, respectively.

Median income level is 500. The lower income group include the median one, and so observations of the low-income group are larger than the high-income group.

[6] Following the suggestion of an anonymous referee, we have introduced this theoretical model to formalize our hypotheses. We are grateful for the referee's constructive comments, which greatly improved the theoretical foundation of this study.

factor $k$. This formulation implies that an individual derives greater utility when their own contribution effectively increases the welfare of the poor relative to the overall social income level.

The second term, $z(tY)$, reflects the satisfaction obtained from the act of giving itself (where $z$ represents the degree of the "warm glow"). The third term, $\ln(Y - tY)$, represents utility derived from one's own disposable income.

The first-order condition (FOC) with respect to $t$ gives the following expression for the optimal tax rate.

$$\frac{dU}{dt} = \frac{-Y}{Y-tY} + \frac{wkY}{Y_p+ktY} + \frac{zY}{tY} = 0 \tag{2}$$

Then, we obtain,

$$F(t;k) = -\frac{1}{1-t} + \frac{wk}{Y_p/Y+kt} + \frac{z}{t} = 0 \tag{3}$$

$$\frac{\partial F}{\partial t} = F_t = -\frac{1}{(1-t)^2} - \frac{wk^2}{(Y_p/Y+kt)^2} - \frac{z}{t^2} < 0 \tag{4}$$

$$\frac{\partial F}{\partial k} = \frac{w(Y_p/Y+kt)-wkt}{(Y_p/Y+kt)^2} = \frac{w(Y_p/Y)}{(Y_p/Y+kt)^2} \tag{5}$$

$$\frac{dt}{dk} = -\frac{\partial F/\partial K}{\partial F/\partial t} \tag{6}$$

From (3) and the implicit function theorem, we have,

$$\frac{dt}{dk} = -\frac{\frac{w(Y_p/Y)}{(Y_p/Y+kt)^2}}{F_t} \tag{7}$$

Its sign depends on $w(Y_p/Y)$ because $F_t$ is always negative. $Y_p/Y$ is constant and so

$\frac{dt^*}{dk} > 0$ if $w > 0$; $\frac{dt^*}{dk} = 0$ if $w = 0$.

This means that if an individual is altruistic ($w > 0$), then a higher efficiency $k$ increases the marginal utility of their contribution to the poor, thereby leading to a higher optimal tax rate. Conversely, if an individual is motivated only by warm glow ($w = 0$, $Z > 0$), the efficiency $k$ disappears from the marginal utility of the tax rate, and thus the optimal tax rate remains invariant to changes in $k$.

In reality, sleep hours are not constant. Thus, respondents were asked about their average wake-up and bedtimes on weekdays. From this data, their sleep hours were calculated. However, some respondents seemed to answer the questions incorrectly. For instance, a sleep time shorter than two or three hours is unrealistic because they would have to take a nap during the day to compensate for the lack of sleep. That is, respondents possibly woke up and went to bed several times a day. We used several sub-samples after excluding excessively short or long sleep hours.

To scrutinize the motivation to pay tax, we set a hypothetical situation. Specifically, by assuming a hypothetical scenario in which an individual's tax contributions are redistributed to low-income groups, this study compares the effects of varying redistribution parameters. To distinguish "pure" and "impure" altruism, it is critical to consider how people change their pro-social behavior if the behavior can more effectively improve another's situation.

First, we examined how an increase in sleep is associated with the allowable tax rate to be spent exclusively on income distribution to the poor. We could not tell whether the allowable tax rate in this question was based on "pure" or "impure" altruism. In the second step, we asked how they would change the allowable tax rate in response to an increase in the contribution paid to improve the situation of the poor.

Key dependent variables were obtained from the following questions under the hypothetical setting:

> *"Q1. Assume that 80 percent of Japan's population earns less than one-fifth of your income. Further, suppose that the tax paid by you goes directly to those with lower incomes. What percentage of your income would you be willing to pay as tax?*
>
> *Please indicate your allowable tax as above. Choose from 1–50 percent."*
>
> *"Q2. Assume that 80 percent of the Japanese population earns less than one-fifth of your income. Further, assume that twice the amount of tax you pay goes to those with lower incomes. Compared to the answer to Q1, what tax burden would you accept in this case?*
>
> *Please choose one of these three choices,*
>
> *(1) Smaller tax burden, (2) Will not change, (3) Greater tax burden."*

In Q1, the allowable tax rate measures the preference for redistribution. To keep the situation equal for high- and low-income people in the real world, we assumed that all respondents had a far higher income than average. Thus, we assumed unrealistically large income inequality. We asked Q2 immediately after Q1 to obtain a proxy variable for assessing whether paying tax was based on pure altruism. If participants intended to improve the conditions of low-income households, they would change their allowable tax rate, assuming that the tax paid would have a two-fold larger effect.

## 2.3  Basic Statistics

In Table 1, we provide definitions of the key variables plus their mean values and standard errors for the whole sample and the sub-samples of high- and low-income households.

<Table 1>

As shown in Table 1, the average *Sleep* is 6.71 for high-income households was slightly shorter than 6.97 for low-income households. In this study, respondents were categorized as high income if their household income was higher than the median (5 million yen) or low income if their household income was equal to or lower than the median. Consistent with this, the mean value of *Income* for the high-income group was 10.4 million yen, three times higher than that for the low-income group (3.34 million yen). The *Allowable tax rate* was 8.65 for high-income households, which is 0.5 percentage points higher than that for low-income households. Similarly, concerning *Increase allowable tax,* share of those who accept greater tax burden are 10.5 and 9.7 percentage points for high- and low- income earners, respectively.

<Fig. 1(a)>

Figure 1(a) illustrates the distribution of sleep hours in the entire sample. Approximately 2 percent sleep 1 or 2 h, whereas 1 percent sleep for 11 h. Therefore, to construct the subsamples used for the estimations, we limited the sample period to 3–10 hours. Other answers were considered outliers, probably due to incorrect answers regarding bedtime and wake-up time. About 33 percent of respondents answered 7 sleep hours, while 23 percent reported six hours.

<Fig. 1(b)>

<Fig. 1(c)>

Both the quantity and quality of sleep may be correlated subjectively. Sleep quality is believed to improve when one goes to bed before midnight (Cellini et al., 2020, 2021; Di Giorgio et al., 2021). Bedtime was used to measure sleep quality.

Figure 1(b) illustrates the distribution of bedtimes, showing that people were divided into two groups: those who went to bed between 21:00 and 24:00 and those who went to bed between 01:00 and 20:00. Hence, in this study, the former was called the high-quality sleep group, and the latter was called the low-quality sleep group. Fig. 1(c) compares the distribution of sleep hours between high- and low-quality sleepers. The highest share was seven sleep hours for the high-quality sleep sample and six sleep hours for the low-quality sleep sample. The proportion of those who were asleep for less than five hours was distinctly larger for the low-quality sleep group than for the high-quality sleep group. Therefore, the higher the quality of sleep, the higher is the quantity of sleep.

<Fig. 2>

Figure 2 shows the distribution of allowable tax rates. Respondents chose allowable tax rates from between 1 percent and 50 percent. With the highest share, 30 percent of people chose 1 percent as the allowable tax rate. Although not reported in the tables and figures, for those who chose 1 percent in Q1, about 40 percent chose "reduce their tax rate" in Q2. This means that they would set their tax rate at 0 percent if that were possible. If there were no obligation, they would not pay any tax to improve the situation of poor people. They are completely individualistic and consider solely their own benefit. About 40 percent of people chose a tax rate between 5 percent and 10 percent. Only about 25 percent of people allowed the tax rate to be over 10 percent. Their responses were reasonably modest even when assuming that income inequality is very large. Fig. 2 indicates that only about 3 percent of people chose a tax rate over 30 percent. They were considered as outliers, and including their answers may have introduced bias into the estimation results. Therefore, a sample of only those who chose 30 percent or below was used for the estimations illustrated in Figures 3 to14. In the appendix, we show the results using a sample that included those who chose tax rates over 30 percent.

<Fig. 3. Mean Allowable Tax Rates in Q1 According to Groups of Sleep Hours.>

In figure 3, the mean allowable tax rates for each group of sleep hours are demonstrated. The sample was limited to 3–10 sleep hours, the same as the sample used for the regression estimations in figures 6, 8, and 9. This is because the samples of less than 3 and more than 10 were very small. A cursory examination of figure 3 indicates that the mean allowable tax rates rose from 3 to 7 hours and subsequently fell, although the differences between them were not statistically significant. The highest allowable tax rate was approximately 8 percent for the group with 7 hours of sleep. As is generally known, the optimal sleep time is around 7 hours (Chaput et al., 2020; Li et al., 2019; Shan et al., 2015; Wu et al., 2018; Yin et al., 2017). Therefore, the optimal amount of sleep is associated with the highest allowable tax rate.

<Fig. 4. Change in Allowable Tax Rate (Share of Responses to Q2).>

Figure 4 shows the respondents' attitudes in response to an increase in their tax contribution. Nearly 65 percent of people did not change their allowable tax rate. Twenty percent reduced their tax rate. One possible interpretation is that they set a target for improvement in the standard of living. To this end, the allowable tax rate should be reduced because the effect of their tax contribution becomes larger. To put it differently, their tax burden should be efficiently reduced for the sake of targeted improvement. Approximately 10 percent of people allowed the tax rate to increase.

<Fig. 5 >

Figure 5 shows how the mean allowable tax rates in Q1 varied according to responses to Q2. The difference in mean values was statistically significant. The figure clearly reveals that respondents whose allowable tax rate was higher were more likely to increase the tax rate in Q2. That is, the gap in allowable tax between the high and low allowable tax rate groups widened if the tax contribution would improve the lot of poor people. This shows that people who answered a higher allowable tax rate were more likely to be purely altruistic.

## 3 Method

To explore how sleep hours are associated with redistribution preferences, we used regression estimations. The baseline estimation function is shown below:

$$Allowable\ Tax_i = \alpha_0 + \alpha_1 Sleep_i + \alpha_2 Sleep_i^2 + X_i'A_i + e_i \qquad (8)$$

First, the dependent variable is *Allowable Tax*. In this case, *Allowable Tax* is censored at the lower bound, 1. Hence, we used a Tobit model for estimation. The expected sign of *Sleep* was positive. $X$ is the vector of control variables that include household income, gender dummy, age, job status dummies, and educational background dummies. $e_i$ is an error term. As can be observed in figure 3, the relationship between sleep hours and allowable tax rate is not linear. Hence, for closer examination in the Tobit estimation, the square of *Sleep* was included.

Using another model, we examined the respondents' answers to Q2:

$$Prob(Z_i = j) = F(\beta_1 Allowable\ Tax_i + \beta_2 Allowable\ Tax_i^2 + \beta_3 Sleep_i + \beta_4 Sleep_i^2 + X_i'B_i) \qquad (9)$$

J=1, 2, 3. *Increase allowable tax* is incorporated as $Z_i$. As shown in table 1, three different outcomes were calculated separately in different estimations. In this case, $Z_i$ is discrete ordered variables, and thus an ordered logit model was used. Different from the Tobit model, the marginal effect can be calculated for three different outcomes. $\alpha$ has 3 different values in an estimation because there are the probabilities of (a) a reduction in the allowable tax rate, (b) no change in the allowable tax rate, or (c) an increase in the allowable tax rate if the setting of Q1 is changed to that of Q2.

In addition to the set of independent variables in specification (8), we incorporated *Allowable Tax* as an independent variable. Whether respondents changed the allowable tax rate dpended on the allowable tax rate level before changing the setting.

As demonstrated in figure 5, the higher the *Allowable Tax*, the more likely respondents would increase their allowable tax rate. However, this tendency could change if other factors were controlled. In the ordered logit model, we scrutinized this tendency by controlling for various factors using a set of control variables.

# 4 Results

As a baseline model, we did not include the square terms of *sleep* and *allowable tax rate*. Table 2 presents the results from a sample of those who chose tax rates over 30 percent. Table 2 also reports the results for all independent variables included in the estimated model. Given a certain income, sleep scarcity may differ depending on employment/retirement status. Furthermore, we reported the interaction terms between *sleep* and job status dummies, where the group of full-time workers was set as a reference. When the interaction terms were not included, significant positive signs of *sleep* were observed when the dependent variables were *allowable tax rate* and *increase allowable tax* in columns (1) and (3), respectively. As for the cross terms, only the positive sign of *Retire ×Sleep* shows statistical significance in column (4), where the dependent variable is *increase allowable tax*. That is, the positive correlation between *sleep* and *increase allowable tax* is stronger for retired workers than for full-time workers.

The marginal effect possibly varied according to the number of sleep hours. To visualize the marginal effect of sleep on the allowable tax rate in the tobit model, Figs. 6(a)–(b) show the marginal effects and 95 percent confidence intervals (CIs) according to each sleep hour. Concerning a change in the allowable tax rate, an ordered logit model was used to investigate *Increase allowable tax,* and the probability of choosing each category was explored. Figure 8 illustrates the marginal effect of sleep, while figure 9 illustrates that of the allowable tax rate. In all estimations, control variables were incorporated as independent variables, as explained in section 3, although these results are not reported.

## 4.1  Results Using the Whole Sample

<Figs. 6(a)–(c)>

Figure 6(a) simply reports the results of the baseline model. However, sleep hours capture only sleep quantity. As shown in Figure (b), the data can be broadly categorized into two groups: those who go to bed between 21:00 and 24:00, and those whose bedtime falls between 01:00 and 20:00. Figure 6(b) shows the results for high-quality sleep hours, while figure 6(c) shows those for low-quality ones.

Figures 6(a)–(c) show that the effect is strongest at 3 hours of sleep and then decreases fairly consistently. At a certain point, the effect becomes negative. As shown in figure 6(a), the marginal effect was almost 0 at 7 hours and then negative at over 8 hours. Figure 6(b) shows that the effect was slightly positive at 7 hours and weakest (around -1.5) at 10 hours. Different from this, as shown in figure 6(c), the effect was slightly negative at 7 hours and weakest (around -2.2) at 10 hours.

<Fig. 7>

Using the results in figures 6(b) and (c), figure 7 illustrates the level of allowable tax at each hour by comparing high- and low-quality sleep. The initial allowable tax was set at the mean allowable tax at three hours. Both showed an inverted U-shape, which is consistent with figure 3. Further, high-quality sleep clearly showed a higher allowable tax rate than low-quality sleep. The peak of high-quality sleep was at 7 hours, at about 12.5 percent, whereas that of low-quality sleep was at 6 hours, at 11.9 percent.

<Fig. 8>

Now, we will explore whether the allowable tax rate increased or decreased when the setting was changed. Figures 8 and 9 demonstrate the effects on the probability of choosing 1 (Reduce) or 3 (Increase). To clearly show the contrast between them, the probability of choosing 2 (Not change) is not reported. Figure 8 illustrates the marginal effect of sleep. The longer the sleep hours, the more likely the allowable tax rate was to increase and less likely to decrease. However, statistical significance was only observed at 7 hours. Therefore, the optimal sleep time for health was significantly correlated with changing the allowable tax rate.

<Fig. 9>

Figure 9 demonstrates that those who allowed a tax rate of 1–15 percent were more likely to increase the allowable tax rate and less likely to reduce it. However, this tendency changed at around 20 percent. Those who allowed a tax over 20 percent were less likely to increase the tax rate and more likely to reduce it. As shown in table 1, the situation was changed so that twice the amount of tax an individual paid went to those with lower incomes. Therefore, their contribution became twice the cost they incurred. Those who had already allowed a sufficient tax rate had an incentive to reduce their cost because their contribution would not change. They were motivated to pursue efficiency, rather than altruism. Most respondents were positive toward increasing the tax rate because most of them chose a tax rate lower than 20 percent (figure 2). Hence, pure altruism motivated most people to allow a higher tax rate.

## 4.2 Comparing Results between High- and Low-Income Households

<Fig. 10(a) and (b)>

<Fig. 11(a) and (b)>

Previous research has demonstrated that preferences for income redistribution differ significantly between high-income and low-income groups (Yamamura 2012). We will now analyze the extent to which allowable tax rates change in relation to each income level. Hence, we divided the sample into high- and low-income household groups. Then, we conducted the same estimations as seen in figures 6–9 to compare the sleep–allowable tax correlation between the groups. Figures 10 and 11 are similar to figure 6. The marginal effect was highest at 3 hours and subsequently decreased. For the high-income group, figure 10 (a) shows that the marginal effect became negative at 9 hours, and the lowest value was about -0.4. Figure 10 (b) indicates that the marginal effect became negative at 7 hours, and the lowest value was -1.8. High-quality sleep clearly showed a higher marginal effect than low-quality sleep. For the low-income group, figure 11(a) shows that the marginal effect became 0 at 7 hours, and the lowest value was -2.0. Figure 11 (b) indicates that the marginal effect was not different from 0 at 6 hours, and the lowest value was about -2.2. The difference was due to differences in sleep quality.

<Fig. 12>

The same as figure 7, using the results of figures 10 and 11, we will illustrate the level of allowable tax according to sleep hours. The mean allowable tax rates for the high- and low-income groups were about 9.3 percent and 7.2 percent based on a sample of those who sleep 3 hours. These values were used as the initial values at 3 hours for each group. As shown in figure 12, the highest level of allowable tax was observed for the high-quality sleep and high-income group. The allowable tax rate hardly declined even if sleep time became longer than 7 hours. In contrast, for the other group, the allowable tax rate decreased at over 7 sleep hours. For the low-income group, the difference between high and low sleep quality was smaller than for the high-income group.

<Fig. 13(a) and (b)>

Figures 13 (a) and (b) show the marginal effect of sleep according to sleep hours for the high- and low-income groups, respectively. We can observe a difference between (a) and (b). However, the effects of sleep hours were not at all different from 0. In the ordered logit model, the allowable tax rate was included as an independent variable. Hence, the statistical insignificance of sleep hours reflects that the effect of sleep hours might be correlated with changes in allowable tax through the channel of increased allowable tax rates.

<Fig. 14 (a) and (b)>

Figures 14(a) and (b) are similar to figure 9. As shown in figure 14(a), for the high-income group, the positive effect of allowable tax on the probability of increasing tax was statistically different from 0 between 1 percent and 18 percent. Further, its effect changed to be negative at around 22 percent, although the negative effect was not statistically different from 0. For the low-income group, the positive effect of the allowable tax rate was statistically different from 0 between 1 percent and 13 percent, and changed to be negative at around 17 percent. Further, at over 20 percent, the negative effect was statistically different from 0. These results indicate that the allowable tax rate was more positively correlated with an increase in the tax rate for the high-income group than the low-income one.

# 5  Discussion

Income redistribution is considered a measure for helping low-income individuals. The following findings have been derived from survey data. We found an inverse U-shaped correlation between sleep hours and the allowable tax rate. Furthermore, there is a positive correlation between sleep hours and the probability of increasing the allowable tax rat. One interpretation of our findings is that people become more altruistic as their sleep approaches seven hours. In other words, when optimal sleep time is achieved, people are more likely to be altruistic toward income redistribution. To realize this virtuous circle, an increase in sleep hours is key.

To assess the preference for redistribution, many previous studies explored how preference for income redistribution is formed (Alesina and Giuliano, 2009; Alesina and La Ferrara, 2005; Corneo and Grüner, 2002; Ohtake and Tomioka, 2004; Sabatini et al., 2020; Yamamura, 2012, 2014). However, this approach does not clarify whether the respondents' views are based on "pure altruism" or "impure altruism" (a "warm glow").

The main contribution of this study is its identification of their motivation in a hypothetical situation about allowable tax. The combined information on sleep hours and responses to the second question enabled us to assess how the quantity of sleep is correlated with redistribution preferences and whether the correlation changes if the setting was changed.

The positive correlation between sleep hours and allowable tax rate strengthened when sleep quality improved. Furthermore, this tendency was clearly observed for the high-income group but not for the low-income group.

The approach employed in this study was very simple because we could not set complicated situations like in laboratory experiments. To scrutinize the preference for redistribution, we should examine how people's decisions change according to the setting. In a laboratory experiment, participants are assigned varying levels of initial income to create inequality, and this information is shared anonymously so that individuals cannot be identified. Each participant is informed only of their own income level, without knowing where they stand relative to the overall distribution. Under these conditions, participants are asked to determine a tax rate for income redistribution that offers no personal financial return. If there are no monetary incentives, the tax rate would theoretically be 0%, unless factors such as altruism or a warm glow effect are present. Experiments conducted under these parameters are useful for validating the findings of this study. Furthermore, it is possible to disentangle pure altruism from the "warm glow" effect by conducting an experiment where the tax paid is increased by a fixed ratio before being distributed to the poor. By introducing this multiplier, we can observe how participants adjust their contributions when the social impact of their giving changes.

An online survey experiment where the provided information changes would be valuable for investigating preferences for redistribution (de Bresser and Knoef, 2022; Kuziemko et al., 2015). Considering differences in preferences between those with and without information about inequality would show the effectiveness of policy and provide information about economic inequality.

The estimation results were obtained based on self-reported variables. So, there may have been some bias due to people not answering sleep-related questions correctly. Further, various factors were not controlled, resulting in endogenous bias; thus, more experiments should be done. Also, information on average bedtime and wake time was obtained, and the difference between the two was calculated as sleep time. Then, bedtime was used to define sleep quality. However, some people may sleep more than once a day. That is, some people take naps to make up for poor sleep at night. Unfortunately, given the data used in this study, we could not scrutinize the effect of multiple daily sleeps.

In the main results, we excluded the sample of those who allowed the tax rate to be over 30 percent because they were considered as outliers. As reported in the Appendix, the outliers drastically changed the results in several estimations, although most results were consistent with the main results. The allowance of a high tax burden appears to have been caused by the lack of appropriate incentives to respond in the simple online survey. Further, the correlation between sleep hours and allowable tax burden might have been over-evaluated. The allowable tax rate would decrease if their monetary rewards were actually determined according to their response.

# 6 Conclusion

This is the first study to indicate how the quantity and quality of sleep correlate with a subjective view of the tax rate. This study contributes to bridge the fields of health and public finance by indicating that sleep quantity and quality are critical when considering tax and income inequality issues. However, causality could not be examined due to data limitations. In future studies, it will be necessary to examine causality using sophisticated methods to provide evidence of potential policy implication. These issues remain to be addressed in future studies.

# 7 Appendix

## 7.1 Ethical considerations

All participants were informed of the aim of the study before starting the survey. All survey participants provided consent to participate in the anonymous online survey via the Internet. Participants responded to the survey questions only after agreeing to participate. Consent for participation in the survey was obtained from all the participants. After starting the survey, the participants had the right to quit at any time. The authors did not obtain any personal information of the participants from the NRC.

The survey design was approved by the Ethics Committee of the Graduate School of Economics, Osaka University (approval no. R51127). Approval was obtained in November 2023, and the survey was conducted in February 2024. Data collection was performed in accordance with the relevant guidelines and regulations. This study did not involve experimental manipulation, there were no foreseeable risks involved, and the questionnaire was anonymous.

The data used in this study were part of long-term panel data on lifestyle attitudes, which included a survey on new coronavirus infections. The application documents for the ethics review clearly stated the purpose of the study as "to accumulate basic knowledge for more desirable policy formulation based on changes in social awareness after the COVID-19 pandemic." Sleep data were collected for this study. Therefore, the intent of this study is consistent with the project's sub-title "Survey on Living Environment during Childhood and Current Life Attitudes" that was included on the documents for ethical approval, although the main title is "A Study on the Influence of the New Type of Coronavirus Infection on Lifestyle Consciousness."

# References

Alesina, A., and Giuliano, P. (2009). *Preferences for Redistribution* (14825; NBER Working Papers).

Alesina, A., and La Ferrara, E. (2005). Preferences for redistribution in the land of opportunities. *Journal of Public Economics*, *89*(5–6), 897–931. https://doi.org/10.1016/j.jpubeco.2004.05.009

Almeida, V., Barrios, S., Christl, M., De Poli, S., Tumino, A., and van der Wielen, W. (2021). The impact of COVID-19 on households´ income in the EU. *Journal of Economic Inequality*, *19*(3), 413–431. https://doi.org/10.1007/s10888-021-09485-8

Anderson, C., and Dickinson, D. (2010). Bargaining and trust: the effects of 36-h total sleep deprivation on socially interactive decisions. *Journal of Sleep Research*, *19*(1-Part-I), 54–63. https://doi.org/10.1111/j.1365-2869.2009.00767.x

Andreoni, J. (1989). Giving with Impure Altruism: Applications to Charity and Ricardian Equivalence. *Journal of Political Economy*, *97*(6), 1447–1458. https://doi.org/10.1086/261662

Andreoni, J. (1990). Impure Altruism and Donations to Public Goods: A Theory of Warm-Glow Giving. *The Economic Journal*, *100*(401), 464. https://doi.org/10.2307/2234133

Ben Simon, E., Vallat, R., Barnes, C. M., and Walker, M. P. (2020). Sleep Loss and the Socio-Emotional Brain. *Trends in Cognitive Sciences*, *24*(6), 435–450. https://doi.org/10.1016/j.tics.2020.02.003

Ben Simon, E., and Walker, M. P. (2018). Sleep loss causes social withdrawal and loneliness. *Nature Communications*, *9*(1). https://doi.org/10.1038/s41467-018-05377-0

Bessone, P., Rao, G., Schilbach, F., Schofield, H., and Toma, M. (2021). The Economic Consequences of Increasing Sleep among the Urban Poor. *Quarterly Journal of Economics*, *136*(3), 1887–1941. https://doi.org/10.1093/qje/qjab013

Buttrick, N. R., and Oishi, S. (2017)."The psychological consequences of inequality." *Social and

*Personality Psychology Compass,* 11(3), e12304

Cellini, N., Canale, N., Mioni, G., and Costa, S. (2020). Changes in sleep pattern, sense of time and digital media use during COVID-19 lockdown in Italy. *Journal of Sleep Research*, *29*(4). https://doi.org/10.1111/jsr.13074

Cellini, N., Conte, F., De Rosa, O., Giganti, F., Malloggi, S., Reyt, M., Guillemin, C., Schmidt, C., Muto, V., and Ficca, G. (2021). Changes in sleep timing and subjective sleep quality during the COVID-19 lockdown in Italy and Belgium: age, gender and working status as modulating factors. *Sleep Medicine*, *77*, 112–119. https://doi.org/10.1016/j.sleep.2020.11.027

Chaput, J. P., Dutil, C., Featherstone, R., Ross, R., Giangregorio, L., Saunders, T. J., Janssen, I., Poitras, V. J., Kho, M. E., Ross-White, A., and Carrier, J. (2020). Sleep duration and health in adults: an overview of systematic reviews. *Applied Physiology, Nutrition, and Metabolism = Physiologie Appliquee, Nutrition et Metabolisme*, *45*(10), S218–S231. https://doi.org/10.1139/apnm-2020-0034

Corneo, G., and Grüner, H. P. (2002). Individual preferences for political redistribution. *Journal of Public Economics*, *83*(1), 83–107. https://doi.org/10.1016/S0047-2727(00)00172-9

Costa-Font, J., Fleche, S., and Pagan, R. (2024). The labour market returns to sleep. *Journal of Health Economics*, *93*, 102840. https://doi.org/10.1016/j.jhealeco.2023.102840

de Bresser, J., and Knoef, M. (2022). Eliciting preferences for income redistribution: A new survey item. *Journal of Public Economics*, *214*, 104724. https://doi.org/10.1016/j.jpubeco.2022.104724

Demichelis, O. P., Grainger, S. A., Burr, L., and Henry, J. D. (2023). Emotion regulation mediates the effects of sleep on stress and aggression. *Journal of Sleep Research*, *32*(3). https://doi.org/10.1111/jsr.13787

Di Giorgio, E., Di Riso, D., Mioni, G., and Cellini, N. (2021). The interplay between mothers' and children behavioral and psychological factors during COVID-19: an Italian study. *European Child and Adolescent Psychiatry*, *30*(9), 1401–1412. https://doi.org/10.1007/s00787-020-01631-3

Dickinson, D. L., and McElroy, T. (2017). Sleep restriction and circadian effects on social decisions. *European Economic Review*, 97, 57–71. https://doi.org/10.1016/j.euroecorev.2017.05.002

Gibson, M., and Shrader, J. (2018). Time use and labor productivity: The returns to sleep. *Review of Economics and Statistics*, 100(5), 783–798. https://doi.org/10.1162/rest_a_00746

Gordon, A. M., and Chen, S. (2013). The Role of Sleep in Interpersonal Conflict: Do Sleepless Nights Mean Worse Fights? Do Sleepless Nights Mean Worse Fights? *Social Psychological and Personality Science,* 5(2), 168-175. https://doi.org/10.1177/1948550613488952

Guadagni, V., Burles, F., Ferrara, M., and Iaria, G. (2014). The effects of sleep deprivation on emotional empathy. *Journal of Sleep Research*, 23(6), 657–663. https://doi.org/10.1111/jsr.12192

Guadagni, V., Burles, F., Valera, S., Hardwicke-Brown, E., Ferrara, M., Campbell, T., and Iaria, G. (2017). The Relationship Between Quality of Sleep and Emotional Empathy. *Journal of Psychophysiology*, 31(4), 158–166. https://doi.org/10.1027/0269-8803/a000177

Holbein, J. B., Schafer, J. P., and Dickinson, D. L. (2019). Insufficient sleep reduces voting and other prosocial behaviours. *Nature Human Behaviour*, 3(5), 492–500. https://doi.org/10.1038/s41562-019-0543-4

Jahrami, H., BaHammam, A. S., Bragazzi, N. L., Saif, Z., Faris, M., and Vitiello, M. V. (2021). Sleep problems during the COVID-19 pandemic by population: a systematic review and meta-analysis. *Journal of Clinical Sleep Medicine*, 17(2), 299–313. https://doi.org/10.5664/jcsm.8930

Kajitani, S. (2021). The return of sleep. *Economics and Human Biology*, 41, 100986. https://doi.org/10.1016/j.ehb.2021.100986

Kamphuis, J., Meerlo, P., Koolhaas, J. M., and Lancel, M. (2012). Poor sleep as a potential causal factor in aggression and violence. *Sleep Medicine*, 13(4), 327–334. https://doi.org/10.1016/j.sleep.2011.12.006

Kuziemko, I., Norton, M. I., Saez, E., and Stantcheva, S. (2015). How Elastic Are Preferences for Redistribution? Evidence from Randomized Survey Experiments. *American Economic Review*, 105(4), 1478–1508. https://doi.org/10.1257/aer.20130360

Li, H., Ren, Y., Wu, Y., and Zhao, X. (2019). Correlation between sleep duration and hypertension: a dose-response meta-analysis. *Journal of Human Hypertension*, *33*(3), 218–228. https://doi.org/10.1038/s41371-018-0135-1

Monaco, K., Olosson, L., and Hentges, J. (2005). Hours of sleep and fatigue in motor carriage. *Contemporary Economic Policy*, *23*(4), 615–624. https://doi.org/10.1093/cep/byi047

Ohtake, F., and Tomioka, J. (2004). Who Supports Redistribution？ *The Japanese Economic Review*, *55*(4), 333–354. https://doi.org/10.1111/j.1468-5876.2004.00318.x

Palomino, J. C., Rodríguez, J. G., and Sebastian, R. (2020). Wage inequality and poverty effects of lockdown and social distancing in Europe. *European Economic Review*, *129*. https://doi.org/10.1016/j.euroecorev.2020.103564

Patel NP, Grandner MA, Xie D, Branas CC, Gooneratne N. (2010) "Sleep disparity" in the population: poor sleep quality is strongly associated with poverty and ethnicity. BMC Public Health, 11(10), 475. doi: 10.1186/1471-2458-10-475.

Sabatini, F., Ventura, M., Yamamura, E., and Zamparelli, L. (2020). Fairness and the Unselfish Demand for Redistribution by Taxpayers and Welfare Recipients. *Southern Economic Journal*, *86*(3). https://doi.org/10.1002/soej.12416

Sabia, J. J., Wang, K., and Cesur, R. (2017). Sleepwalking through school: new evidence on sleep and academic achievement. *Contemporary Economic Policy*, *35*(2), 331–344. https://doi.org/10.1111/coep.12193

Saghir, Z., Syeda, J. N., Muhammad, A. S., and Balla Abdalla, T. H. (2018). The Amygdala, Sleep Debt, Sleep Deprivation, and the Emotion of Anger: A Possible Connection?. Cureus 10(7): e2912. doi:10.7759/cureus.2912

Shan, Z., Ma, H., Xie, M., Yan, P., Guo, Y., Bao, W., Rong, Y., Jackson, C. L., Hu, F. B., and Liu, L. (2015). Sleep Duration and Risk of Type 2 Diabetes: A Meta-analysis of Prospective Studies. *Diabetes Care*, *38*(3), 529–537. https://doi.org/10.2337/dc14-2073

Simon, E. Ben, Vallat, R., Rossi, A., and Walker, M. P. (2022). Sleep loss leads to the withdrawal of human helping across individuals, groups, and largescale societies. *PLoS Biology*, *20*(8). https://doi.org/10.1371/journal.pbio.3001733

Studler, M., Gianotti, L. R. R., Lobmaier, J., Maric, A., and Knoch, D. (2024). Human prosocial preferences are related to slow-wave activity in sleep. *The Journal of Neuroscience*, e0885232024. https://doi.org/10.1523/jneurosci.0885-23.2024

Wang, B., Isensee, C., Becker, A., Wong, J., Eastwood, P. R., Huang, R. C., Runions, K. C., Stewart, R. M., Meyer, T., G. Brüni, L., Zepf, F. D., and Rothenberger, A. (2016). Developmental trajectories of sleep problems from childhood to adolescence both predict and are predicted by emotional and behavioral problems. *Frontiers in Psychology*, *7*(DEC). https://doi.org/10.3389/fpsyg.2016.01874

Wu, L., Sun, D., and Tan, Y. (2018). A systematic review and dose-response meta-analysis of sleep duration and the occurrence of cognitive disorders. *Sleep and Breathing*, *22*(3), 805–814. https://doi.org/10.1007/s11325-017-1527-0

Wyss, A. M., and Knoch, D. (2022). Neuroscientific approaches to study prosociality. *Current Opinion in Psychology*, *44*, 38–43. https://doi.org/10.1016/j.copsyc.2021.08.028

Yamamura, E. (2012). Social capital, household income, and preferences for income redistribution. *European Journal of Political Economy*, *28*(4). https://doi.org/10.1016/j.ejpoleco.2012.05.010

Yamamura, E. (2014). Trust in government and its effect on preferences for income redistribution and perceived tax burden. *Economics of Governance*, *15*(1). https://doi.org/10.1007/s10101-013-0134-1

Yamamura, E. (2015). Effects of Siblings and Birth Order on Income Redistribution Preferences: Evidence Based on Japanese General Social Survey. *Social Indicators Research*, *121*(2), 589–606. https://doi.org/10.1007/s11205-014-0649-z

Yamamura, E. (2024). Grandchildren and views about consumption tax . *Review of Behavioral Economics*.

Yin, J., Jin, X., Shan, Z., Li, S., Huang, H., Li, P., Peng, X., Peng, Z., Yu, K., Bao, W., Yang, W., Chen, X., and Liu, L. (2017). Relationship of Sleep Duration With All-Cause Mortality and Cardiovascular Events: A Systematic Review and Dose-Response Meta-Analysis of Prospective Cohort Studies. *Journal of the American Heart Association*, *6*(9). https://doi.org/10.1161/JAHA.117.005947

**Table 1: Definitions of Key Variables and Their Basic Statistics**

|  |  | Whole | | High income | | Low income | |
|---|---|---|---|---|---|---|---|
| Variable | Definition | Mean | s.d. | Mean | s.d. | Mean | s.d. |
| *Sleep* | Average duration of sleep time on weekdays. | 6.85 | 1.21 | 6.71 | 1.13 | 6.97 | 1.27 |
| *Allowable Tax* | Q1. Assume that 80 percent of Japan's population earns less than one-fifth of your income. Further, suppose that the tax paid by you goes directly to those with lower incomes. What percentage of your income would you be willing to pay as tax? Please indicate your allowable tax as above. Choose from 1–50 percent. | 8.35 | 10.5 | 8.65 | 10.2 | 8.15 | 10.8 |
| *Increase Allowable Tax* | Q2. Assume that 80 percent of the Japanese population earns less than one-fifth of your income. Further, suppose that twice the amount of tax you pay goes to those with lower incomes. Compared to the answer to Q1, what tax burden would you accept in this case? Please choose one from these three choices, 1. Smaller tax burden, 2. Will not change, 3. Greater tax burden. The values given by the respondents were used for calculating means and standard deviations. | 1: 20.0 percent  2: 69.3 percent  3: 10.7 percent | | 1: 20.2 percent  2: 69.3 percent  3: 10.5 percent | | 1: 21.5 percent  2: 69.7 percent  3: 9.7 percent | |
| *Income* | Household income (million yen). | 6.62 | 4.64 | 10.4 | 4.11 | 3.34 | 1.55 |
| *Female* | 1 if the respondent is female, 0 if otherwise. | 0.47 | 0.49 | 0.47 | 0.47 | 0.48 | 0.49 |
| *Age* | Respondent's age | 47.5 | 13.5 | 48.2 | 11.9 | 46.8 | 14.1 |
| *Schooling* | Schooling years | 14.8 | 1.93 | 14.7 | 1.93 | 14.7 | 1.89 |
| *Retire* | It is 1 if a respondent is retired, and 0 otherwise. | 0.09 | 0.28 | 0.02 | 0.15 | 0.15 | 0.35 |
| *Part~time* | It is 1 if a respondent is a part-time worker, and 0 otherwise. | 0.13 | 0.33 | 0.10 | 0.29 | 0.16 | 0.37 |
| *Student* | It is 1 if a respondent is a Student, and 0 otherwise. | 0.03 | 0.16 | 0.02 | 0.14 | 0.03 | 0.17 |
| *Married* | It is 1 if a respondent is married, and 0 otherwise. | 0.43 | 0.49 | 0.26 | 0.44 | 0.56 | 0.49 |

Note. Observations for the whole, high-income, and low-income groups were 3,641, 1,692, and 1,949, respectively.

The sample size was limited to the estimations shown in figures 1 and 2. After excluding outliers of sleep time below 3 and over 10 hours and allowable tax over 30 percent in the sample used for figures 3–14, observations for whole, high income, and low income were 3,476, 1,625, and 1,851, respectively.

Instead of mean values and standard deviations, the allowable tax values indicate the percentage of choices that the respondents made.

**Table 2. Estimations with Interaction Terms Between Sleep Hours and Job Status Dummies**

|  | *Dep Var= Allowable tax rate* **Tobit model** |  | *Dep Var= Increase Allowable tax* **Ordered Logit model** |  |
| --- | --- | --- | --- | --- |
|  | *(1)* | *(2)* | *(3)* | *(4)* |
| *Sleep* | 0.495*** | 0.432** | 0.079*** | 0.037 |
|  | (0.189) | (0.219) | (0.029) | (0.034) |
| *Allowable tax rate* |  |  | 0.030*** | 0.030*** |
|  |  |  | (0.003) | (0.003) |
| *Ln(Income)* | 1.041*** | 1.045*** | 0.114** | 0.114** |
|  | (0.304) | (0.304) | (0.047) | (0.047) |
| *Retire* | 2.683*** | −1.448 | 0.041 | −1.943*** |
|  | (0.868) | (4.913) | (0.137) | (0.747) |
| *Retire ×Sleep* |  | 0.573 |  | 0.276*** |
|  |  | (0.669) |  | (0.102) |
| *Part-time* | −0.579 | −1.950 | −0.066 | −0.555 |
|  | (0.699) | (3.902) | (0.107) | (0.598) |
| *Part-time ×Sleep* |  | 0.198 |  | 0.071 |
|  |  | (0.552) |  | (0.084) |
| *Student* | 4.513*** | 9.127*** | −0.026 | −1.697 |
|  | (1.510) | (9.352) | (0.242) | (1.431) |
| *Student ×Sleep* |  | −0.628 |  | 0.234 |
|  |  | (1.270) |  | (0.196) |
| *Schooling* | 0.040 | 0.040 | −0.005 | −0.005 |
|  | (0.130) | (0.130) | (0.020) | (0.020) |
| *Female* | −2.595*** | −2.582*** | −0.141* | −0.133* |
|  | (0.463) | (0.463) | (0.073) | (0.073) |
| *Ln(Age)* | 5.047*** | 5.060*** | −0.344** | −0.340** |
|  | (0.908) | (0.908) | (0.144) | (0.144) |
| *Married* | 0.166 | 0.166 | 0.027 | 0.027 |
|  | (0.554) | (0.554) | (0.087) | (0.087) |
| likelihood ratio chi-square | 108 | 109 | 105 | 114 |
| Observations | 3,641 | 3,641 | 3,641 | 3,641 |
| Left-Censored | 957 | 957 |  |  |

Note: ***, **, and * denote statistical significance at the 1 percent, 5 percent, and 10 percent levels, respectively. Numbers are coefficients, whereas numbers within parentheses are robust standard errors.

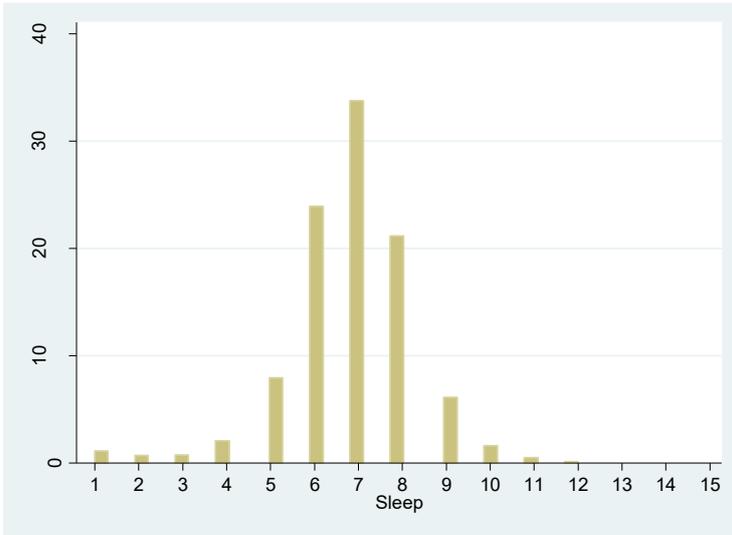

Fig. 1（a）. Distribution of Sleep Hours

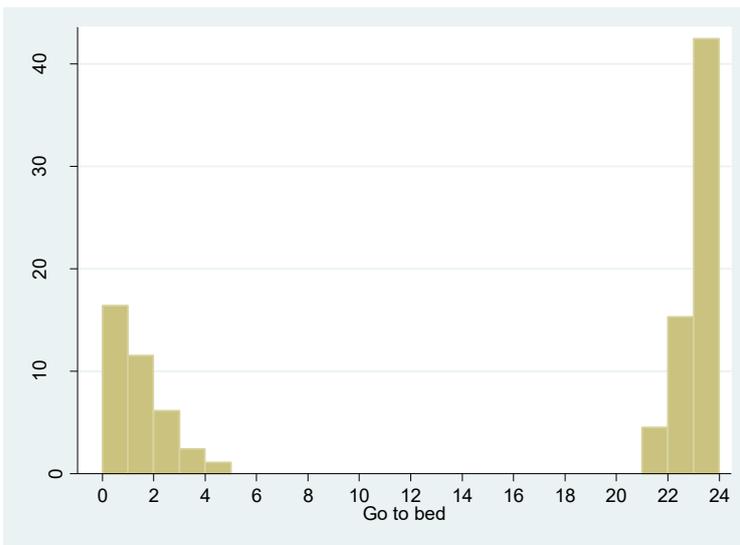

Fig. 1(b). Distribution of Bedtimes

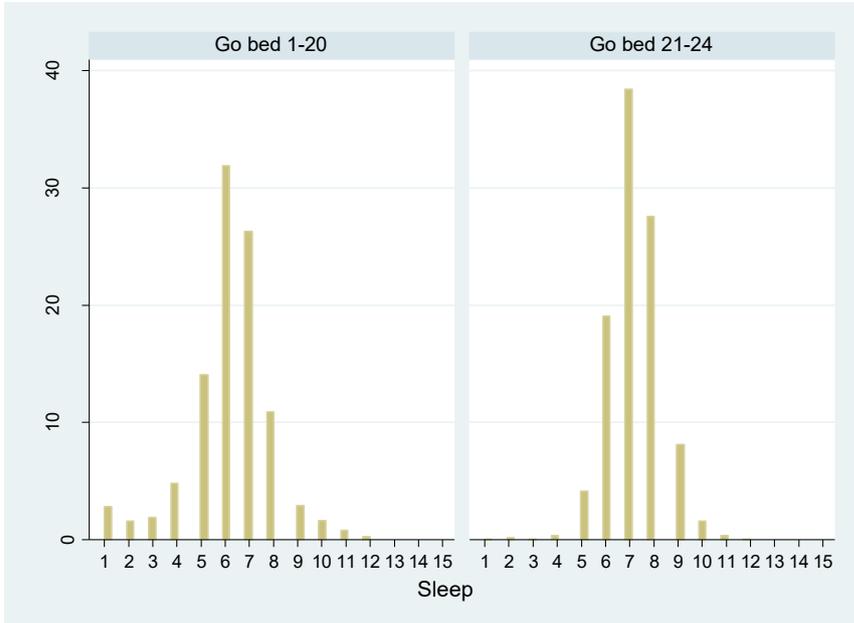

Fig. 1(c). Comparing Distribution of Sleep Hours Between High- and Low-Quality Sleep

Note. 3,731 observations.

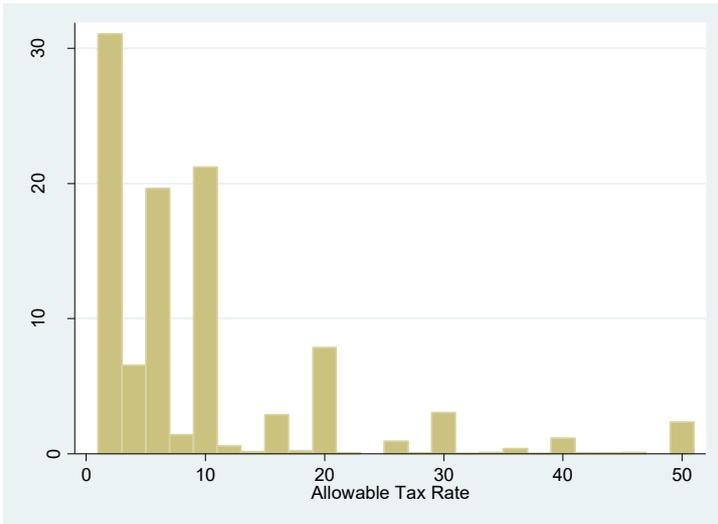

Fig. 2. Distribution of Allowable Tax Rate For Q1

Note. 3,731 observations.

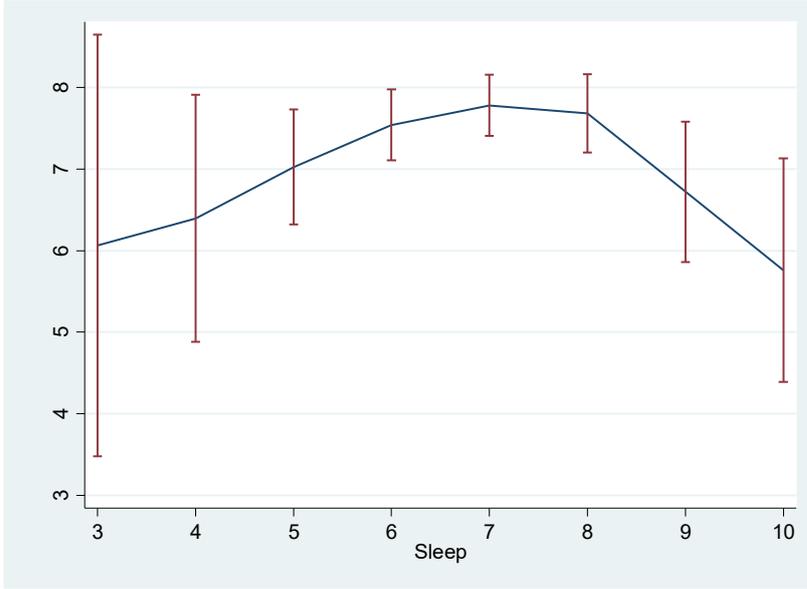

Fig. 3. Mean Allowable Tax Rates in Q1 According to Groups of *Sleep*

Note. A 95 % Confidence Interval was illustrated. 3,467 observations.

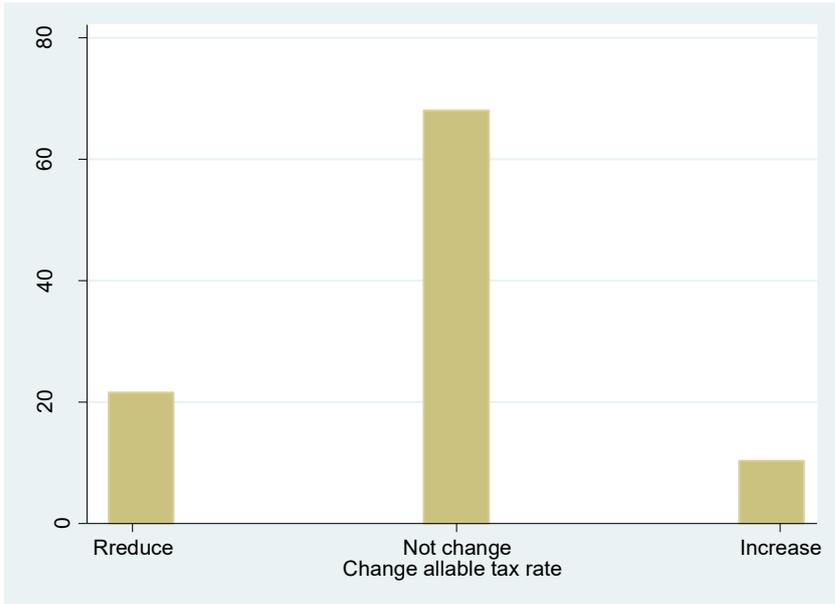

Fig. 4. Change in Allowable Tax Rate (Share of Responses to Q2)

Note. 3,467 observations.

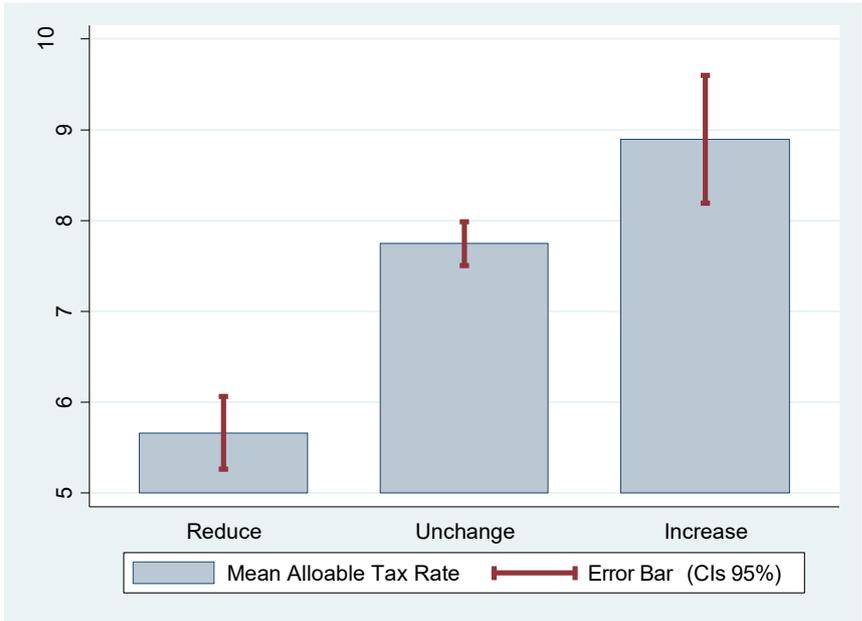

Fig. 5. Mean Allowable Tax Rate for Q1 According to Groups of Responses to Q2

Note. A 95 % Confidence Interval was indicated. 3,467 observations.

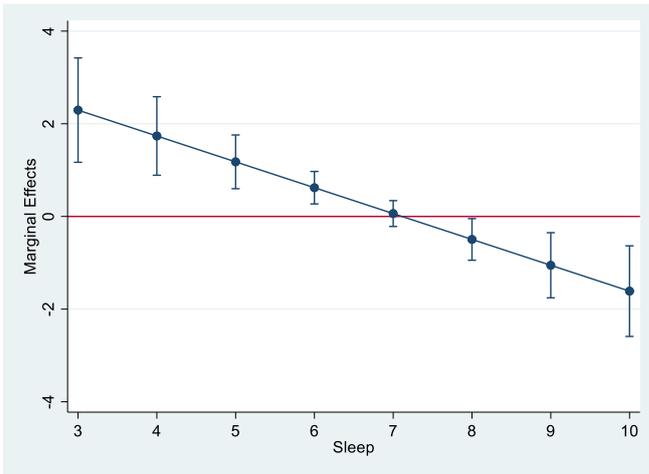

Fig. 6(a) Average Marginal Effects of *Sleep* on *Allowable Tax* with a 95 % Confidence

Interval (Full Sample)

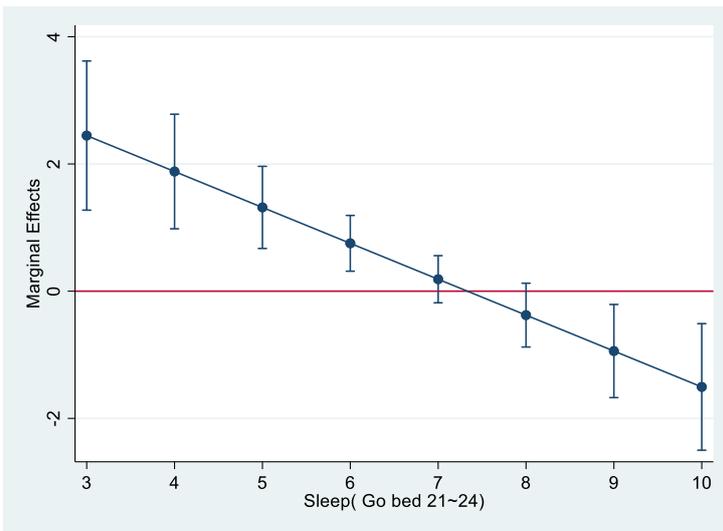

Fig. 6(b) Average Marginal Effects of *Sleep* on *Allowable Tax* with a 95 % Confidence

Interval: High-Quality Sleep

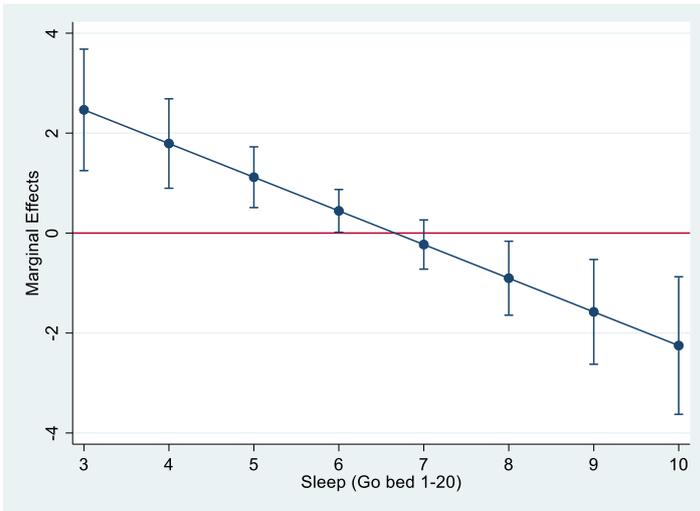

Fig. 6(c) Average Marginal Effects of *Sleep* on *Allowable Tax* with A 95 % Confidence

Interval: Low-Quality Sleep

Note. A 95 % Confidence Interval was obtained using robust standard errors clustered in residential prefectures. Various control variables are included in all columns, such as age, educational background, marital status, and occupational dummies. 3,467 observations.

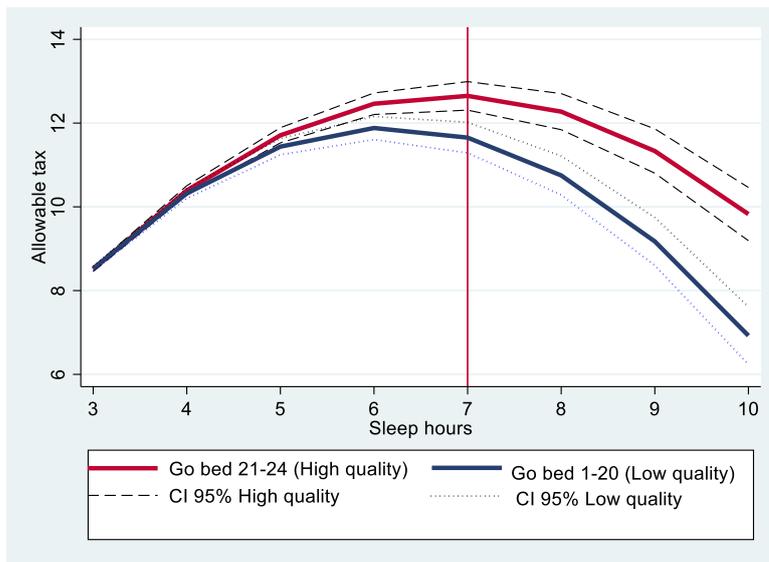

Fig. 7. Allowable Tax and Sleep Hours According to Time to Go to Bed with a 95 % Confidence Interval

Note. A 95 % Confidence Interval was obtained using robust standard errors clustered in residential prefectures. Various control variables are included in all columns, such as age, educational background, marital status, and occupational dummies. 3,467 observations.

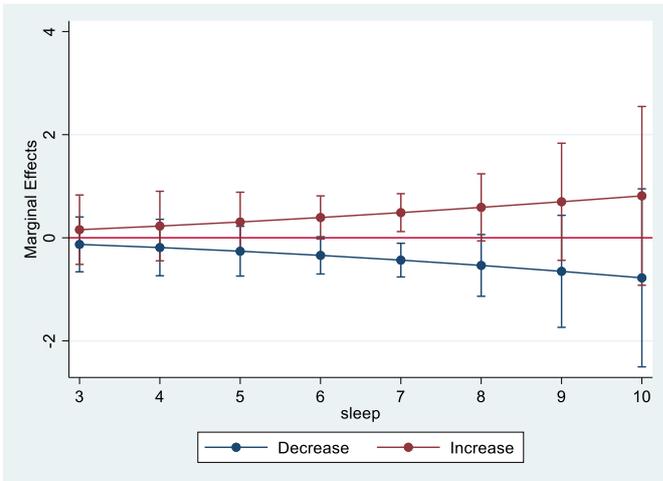

Fig. 8. Average Marginal Effects of *Sleep* on *Increase Allowable Tax* with a 95 % Confidence Interval

Note. A 95 % Confidence Interval was obtained using robust standard errors clustered in residential prefectures. Various control variables are included in all columns, such as age, educational background, marital status, and occupational dummies. 3,467 observations.

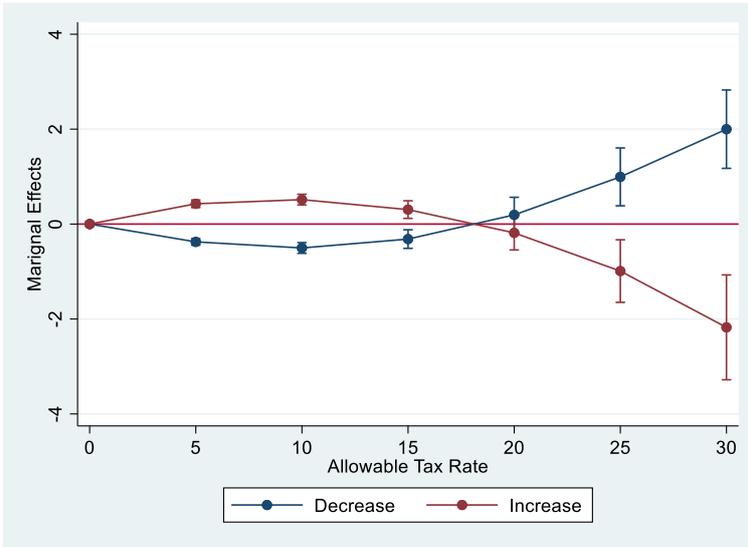

Fig. 9. Average Marginal Effects of *Allowable Tax* on *Probability of Increasing Allowable*

with a 95 % Confidence Interval

Note. A 95 % Confidence Interval was obtained using robust standard errors clustered in residential prefectures. Various control variables are included in all columns, such as age, educational background, marital status, and occupational dummies. 3,467 observations.

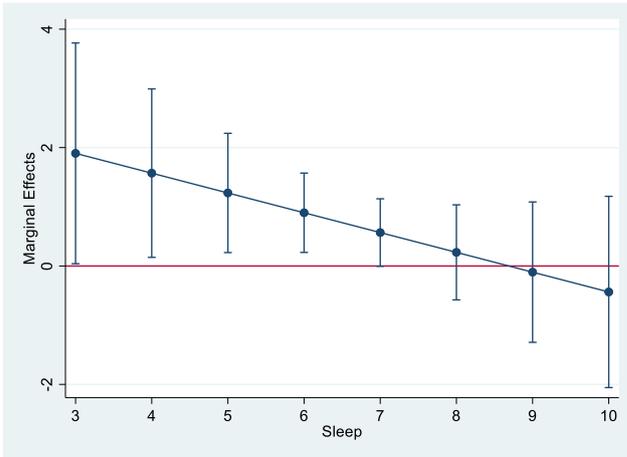

Fig. 10(a). Average Marginal Effects of *Sleep* on *Allowable Tax* with A 95 % Confidence Interval: High-Quality Sleep for High-Income Group

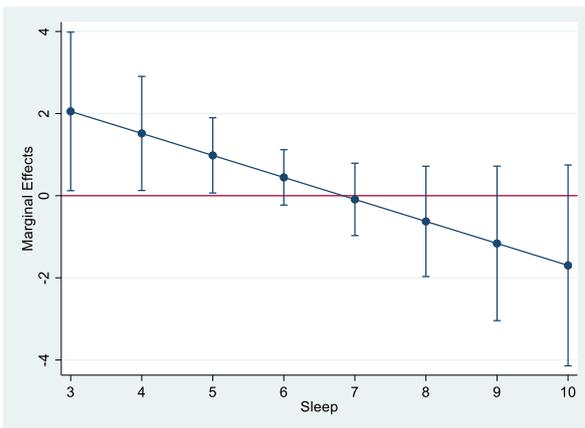

Fig. 10(b) Average Marginal Effects of *Sleep* on *Allowable Tax* with A 95 % Confidence Interval: Low-Quality Sleep for High-Income Group

Note. A 95 % Confidence Interval was obtained using robust standard errors clustered on residential prefectures. Various control variables are included in all columns, such as ages, educational background dummies, marital status dummy, and occupational dummies. 1,625 observations.

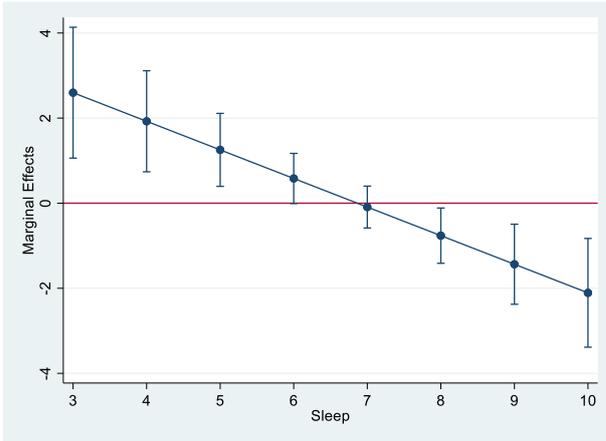

Fig. 11(a) Average Marginal Effects of *Sleep* on *Allowable Tax* with A 95 % Confidence Interval: High-Quality Sleep for Low-Income Group

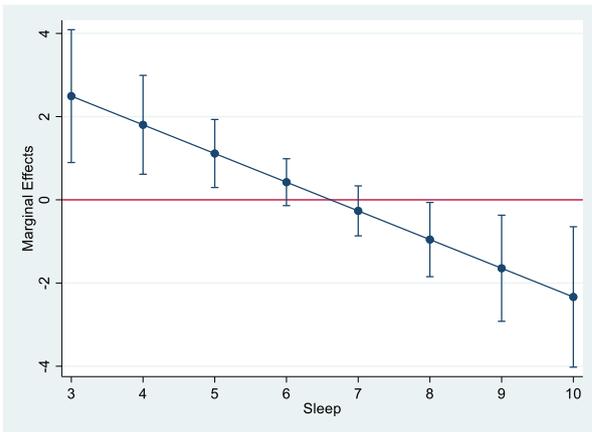

Fig. 11(b) Average Marginal Effects of *Sleep* on *Allowable Tax* with A 95 % Confidence Interval: Low-Quality Sleep for Low-Income Group

Note. A 95 % Confidence Interval was obtained using robust standard errors clustered in residential prefectures. Various control variables are included in all columns, such as age, educational background, marital status, and occupational dummies. 1,851 observations.

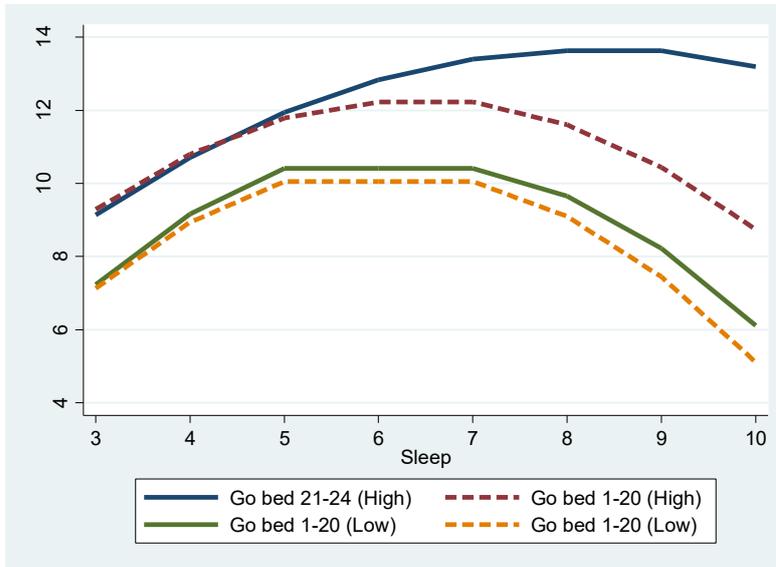

Fig. 12. Relationship Between Estimated Allowable Tax and Sleep Hours: Sub-Sample Results of Four Groups According to Sleep Quality and Income Level

Note. 1,625 and 1,851 observations for the high- and low-income groups, respectively.

2,194 and 1,281 observations for the high- and low-quality sleep groups, respectively.

1,054 observations for high-quality and high-income group.

571 observations for low-quality and high-income group.

1,140 observations for high-quality and low-income group.

711 observations for low-quality and low-income group.

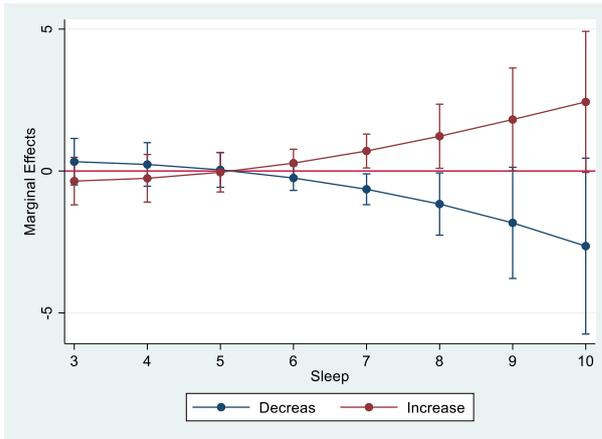

Fig. 13(a) Average Marginal Effects of *Sleep* on *Increase Allowable* with a 95 % Confidence Interval: Results of High-Income Group

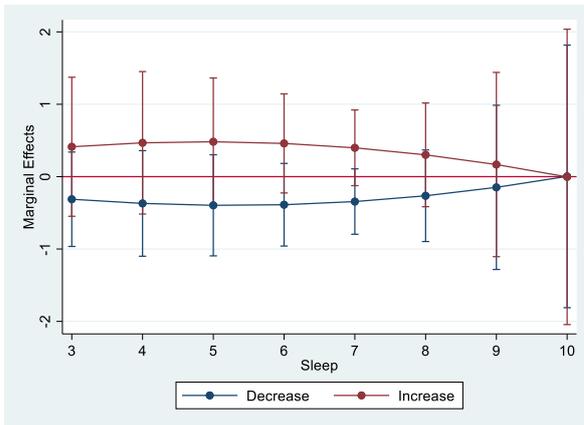

Fig. 13(b) Average Marginal Effects of *Sleep* on *Increase Allowable* with a 95 % Confidence Interval: Results Of Low-Income Group

Note. A 95 % Confidence Interval was obtained using robust standard errors clustered in residential prefectures. Various control variables are included in all columns, such as age, educational background, marital status, and occupational dummies. 1,625 and 1,851 observations for high- and low-income groups, respectively.

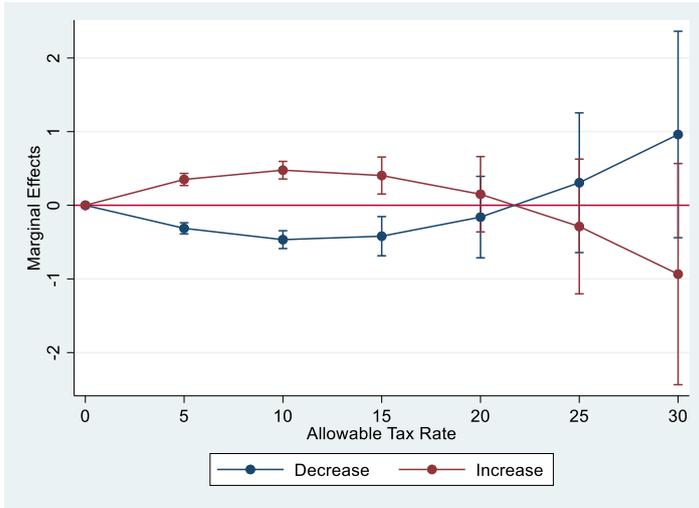

Fig. 14(a) Average Marginal Effects of *Allowable Tax* on *Probability of Increasing Allowable Tax* with a 95 % Confidence Interval: Results of High-Income Sample

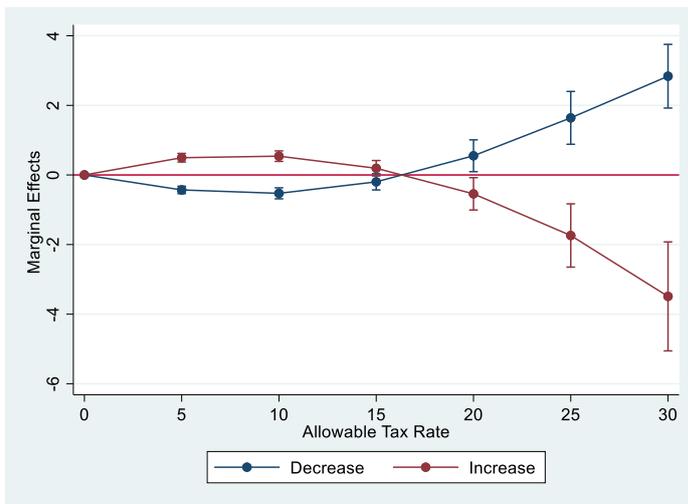

Fig. 14(b). Average Marginal Effects of *Allowable Tax* on *Probability of Increasing Allowable Tax* with a 95 % Confidence Interval: Results of Low-Income Sample

Note. A 95 % Confidence Interval was obtained using robust standard errors clustered on residential prefectures. Various control variables are included in all columns, such as ages, educational background dummies, marital status dummy, and occupational dummies. 1,625 and 1,851 observations for high- and low-income samples, respectively.